\documentclass[%
 reprint,
superscriptaddress,
 amsmath,amssymb,
 aps,
pra,
]{revtex4-2}

\usepackage{graphicx,float}
\usepackage{dcolumn}
\usepackage{bm, hyperref}

\begin{document}

\preprint{APS/123-QED}

\title{Robustness against disorder in topological fibre lasers with explicitly broken \mbox{$\mathcal{PT}$ symmetry}}

\author{Brook Salter}
\affiliation{%
 Department of Physics, University of Bath, Claverton Down, Bath, BA2 7AY, United Kingdom
}

\author{Nathan Roberts}
\affiliation{
National Research Council of Canada, 100 Sussex Drive, Ottawa, Ontario, K1A 0R6, Canada}

\author{Habib Rostami}

\affiliation{%
 Department of Physics, University of Bath, Claverton Down, Bath, BA2 7AY, United Kingdom
}

\author{Peter J. Mosley}
\affiliation{%
 Department of Physics, University of Bath, Claverton Down, Bath, BA2 7AY, United Kingdom \\
 Centre for Photonics, University of Bath, Claverton Down, Bath, BA2 7AY, United Kingdom \\
 ORCA Computing Ltd, 30 Eastbourne Terrace, London W2 6LA, UK
}

\author{Anton Souslov}
\email{as3546@cam.ac.uk}
\affiliation{
    TCM Group, Cavendish Laboratory, J.J. Thomson Avenue, Cambridge, CB3 0US, United Kingdom
}%

\date{\today}% It is always \today, today,

\begin{abstract}
\noindent
Fibre lasers realise a large gain medium in a compactly coiled fibre. Disorder due to fabrication can negatively impact the stability of their lasing modes, especially in multi-core fibres. Recently, topological fibres (without gain) have been experimentally demonstrated to be robust against fabrication disorder, but topological fibre lasers have not yet been designed or modelled.
Here, we use a combination of mode-coupling theory and finite-element simulations to design and model a topological laser based on a non-Hermitian Su-Schrieffer-Heeger (SSH) chain embedded in a photonic crystal fibre. Our design is based on a winding-number invariant in combination with a $\mathcal{PT}$-symmetric SSH bulk. We show that the topological boundary mode is selectively amplified when extra gain is added at the topological interface. Even with nonlinearity added through saturable gain, the lasing supermode retains its robustness against disorder.
We present a realistic design for a topologically robust fibre laser using readily available stack-and-draw methods with doped cores. This work establishes a new approach for imbuing non-Hermitian photonic systems with topological protection, with technological implications towards generating robust quantum and classical signals. 
\end{abstract}

%\keywords{Suggested keywords} Non-Hermitian Topology, Fibre Optics, Nonlinear Optics.

\maketitle

\begin{figure*}[t!]
    \includegraphics[width=1\textwidth]{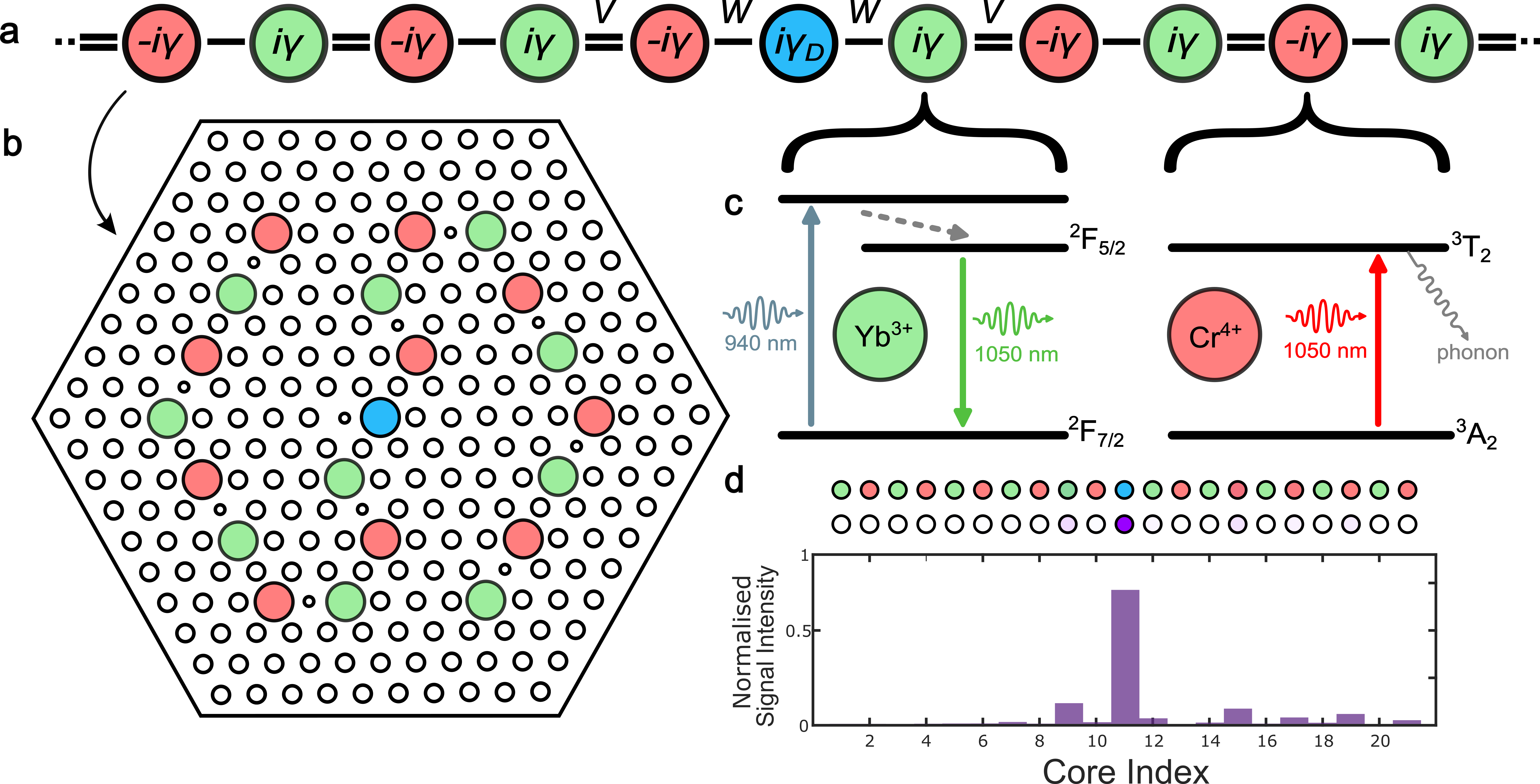}
    
    \caption{\textbf{Proposed fibre design.}\quad \textbf{a} \quad Schematic of the non-Hermitian Su-Schrieffer-Heeger (SSH) chain with a central defect at the interface. Weak and strong couplings are labelled with single and double bonds, respectively. Green and red regions correspond to net optical gain (via stimulated emission from pumped $\mathrm{Yb}^\mathrm{3+}$ ion dopants) or loss (via resonant $\mathrm{Cr}^\mathrm{4+}$ absorption with phononic dissipation at the same operating wavelength) with a magnitude of $\pm i\gamma$, respectively. The central, blue, gain site is variable with magnitude $\gamma_D \geq \gamma.$ \textbf{b}\quad Schematic of a PCF fibre design containing a coiled replica of the SSH chain. Red, green, and blue labellings within signify gain/loss/defect core, respectively. Hollow rings represent 2 sizes of airholes within a global, silica background. Coloured regions are also silica glass, doped with their respective concentration of amplifying and attenuating dopants. \textbf{c} \quad Three-level diagrams of photonic interaction with Yb and Cr dopants, used in the gain and loss sites, respectively. \textbf{d} \quad Heat-map and intensity plot of the signal steady state across all the cores of the chain.}  
    \label{F:1}
\end{figure*}

\section{Introduction}\label{S:Intro}

Topological photonic systems, designed with appropriate symmetries, enable the robust transport of light that is intrinsically protected against fabrication disorder. Drawing inspiration from the topological insulators of condensed matter physics, photonic lattices may be engineered to included topological interfaces harbouring robust boundary states. Non-Hermitian topology has been demonstrated in several physical systems, including but not limited to: active matter \cite{Active1, Active2, Active3, Active4}, superconductivity \cite{Super1, Super2, Super3}, acoustics \cite{AcouMain,Acou1,Acou2}, and photonics. More specifically within the photonic domain, topological lasing has been achieved in polariton arrays \cite{Poln1, Poln2}, solitons, \cite{Soli1, Soli2}, and microring resonator arrays \cite{LaserExp,LaserTheory, Microring1}.\\

Despite the ubiquitous nature of non-Hermitian topology, these non-Hermitian boundary states have yet to be designed in optical fibre. Such a design could have technological implications, for example in enabling higher output power for fibre lasers.  This is due to the recent development of multicore geometries, which spread spread optical gain over many light-guiding regions within a single fibre strand~\cite{Multilaser2, Multilaser1, Multilaser3}, to overcome the challenges in power dissipation of single-core fibres~\cite{Limit1}. This platform shift introduces its own challenges, including increased susceptibility to fabrication disorder, due to the typical requirement for coherent recombination of light that has been amplified in separate regions of the cross-section. Here, we consider whether topology can be exploited in multicore fibre to reduce sensitivity to these errors. Previously, Hermitian optical systems have reportedly displayed robust topological protection in cases of both broken chiral symmetry~\cite{BathSSH} and broken time-reversal symmetry~\cite{BathTwist}. Both were bespoke fibres displaying robust topological boundary modes within a single, compact strand. 

In this work, we investigate the non-Hermitian extension of these topological fibres, thereby extending the reach of topological lasing into a platform central to modern photonic networks. Fibre geometry offers long propagation lengths for the accumulation of non-Hermitian effects, allowing flexible gain-loss engineering with low effective gain per unit length, and a compact architecture that can host microscopic lattice structures within a single strand. As well as serving as a host for protected boundary transport, fibre also provides a natural setting for topological lasing, where lasing action can be pinned to robust boundary modes immune to disorder and mode competition. Such a topological fibre laser would combine the resilience provided by topology with the scalability and cost-effectiveness of fibre technology. This has potential applications for the development of practical, high-performance sources for communications, sensing and quantum technology. \\

We draw inspiration from parity-time ($\mathcal{PT}$)-symmetric systems, while simultaneously considering the role of topology throughout the design of a coupled, multicore fibre~\cite{PT1,PT2}. We show that the explicit breaking of this symmetry in conjunction with non-Hermitian topology allows for the preferential amplification of Su-Schrieffer-Heeger (SSH) boundary modes in the saturable, nonlinear regime.

\section{Theoretical Model}\label{S:2}

There are two key ingredients to the presented system that are worth discussing in detail. The proposed fibre hosts a topologically-protected mode by consequence of the non-Hermitian Su-Schrieffer-Heeger (SSH) model. Preferential lasing of this mode and suppression of other 'bulk' modes is then achieved by $\mathcal{PT}$-symmetry breaking. Below is a brief discussion of each.

\subsection{$\mathcal{PT}$-Symmetric Systems}

A key feature of many non-Hermitian systems is $\mathcal{PT}$ symmetry, which enables real eigenspectra despite non-Hermiticity~\cite{PTreal}. A Hamiltonian, $\mathcal{H} = T + V$, is
$\mathcal{PT}$-symmetric if it commutes with both the parity and time-reversal operators: $[\mathcal{H},\mathcal{PT}] = 0$~\cite{PT2}. Since the kinetic term, $T$, commutes with both of these operators, the necessary condition depends on the potential, $V$. These quantum-mechanical concepts can be utilized in photonics through the formal mapping between the Schrödinger equation and the Maxwell equations in the paraxial approximations~\cite{Maxwell}. The evolution of a slowly-varying electric field propagating along the longitudinal $z$-direction of the fibre satisfies the equation
\begin{align}
    i\partial_z \Psi = - \frac{1}{2\beta}\nabla_t^2\Psi - \frac{k_0}{n_0}\Delta n(x,y) \Psi,
\end{align}
which is reminiscent of the Schrödinger equation, but with $z$-propagation along the fibre taking the role of time evolution. By this mapping, time-reversal $\mathcal{T}$ is analogous to the reversal of the direction of propagation, i.e., $z \to - z$. Here, $\Psi$ is the field envelope, $\beta = k_0n_0$, and $\Delta n(x,y) = n(x,y) - n_0$ is the refractive index contrast acting as an effective potential, $V(x,y)$. Taking $V$ to be a one-dimensional potential parametrised by position $x$, $\mathcal{PT}$-symmetry imposes, 
\begin{align}
V(x) & = V^*(-x),\\
\textrm{with}\quad V(x) &= k_0(n_R(x) + i n_I(x)),
\end{align}
where the refractive index, which plays the role of an effective potential, has been decomposed into real part $n_R(x)$ and imaginary part $n_I(-x)$.
This condition is satisfied when $n_R(x)$ is even, $n_R(x) = n_R(-x)$, and $n_I(-x)$ (associated with gain and loss) is odd, $n_I(x) = -n_I(-x)$. Therefore $\mathcal{PT}$-symmetry requires a balanced gain-loss profile. \\

Eigenvalues in coupled systems, respecting $\mathcal{PT}$-symmetry, typically take the form,
\begin{align}
    E_{\pm} \propto \pm\sqrt{\kappa^2 - \gamma^2},
\end{align}
where $\gamma$ is the gain parameter for the system. The square root partitions the parameter space into two distinct regimes,

\begin{itemize}
    \item $\kappa > \gamma$ (symmetry preserving): eigenvalues all real
    \item $\kappa < \gamma$ (symmetry breaking): eigenvalues become purely imaginary, appearing in complex-conjugate pairs.
\end{itemize}

We will see that by explicitly breaking $\mathcal{PT}$-symmetry, spectral changes occur that allow for the preferential amplification of a single mode. By then tuning the lattice design such that this mode sits in a topological band gap, we can achieve selective topological mode amplification. 

\begin{figure*}[t!]
    \includegraphics[width=1\textwidth]{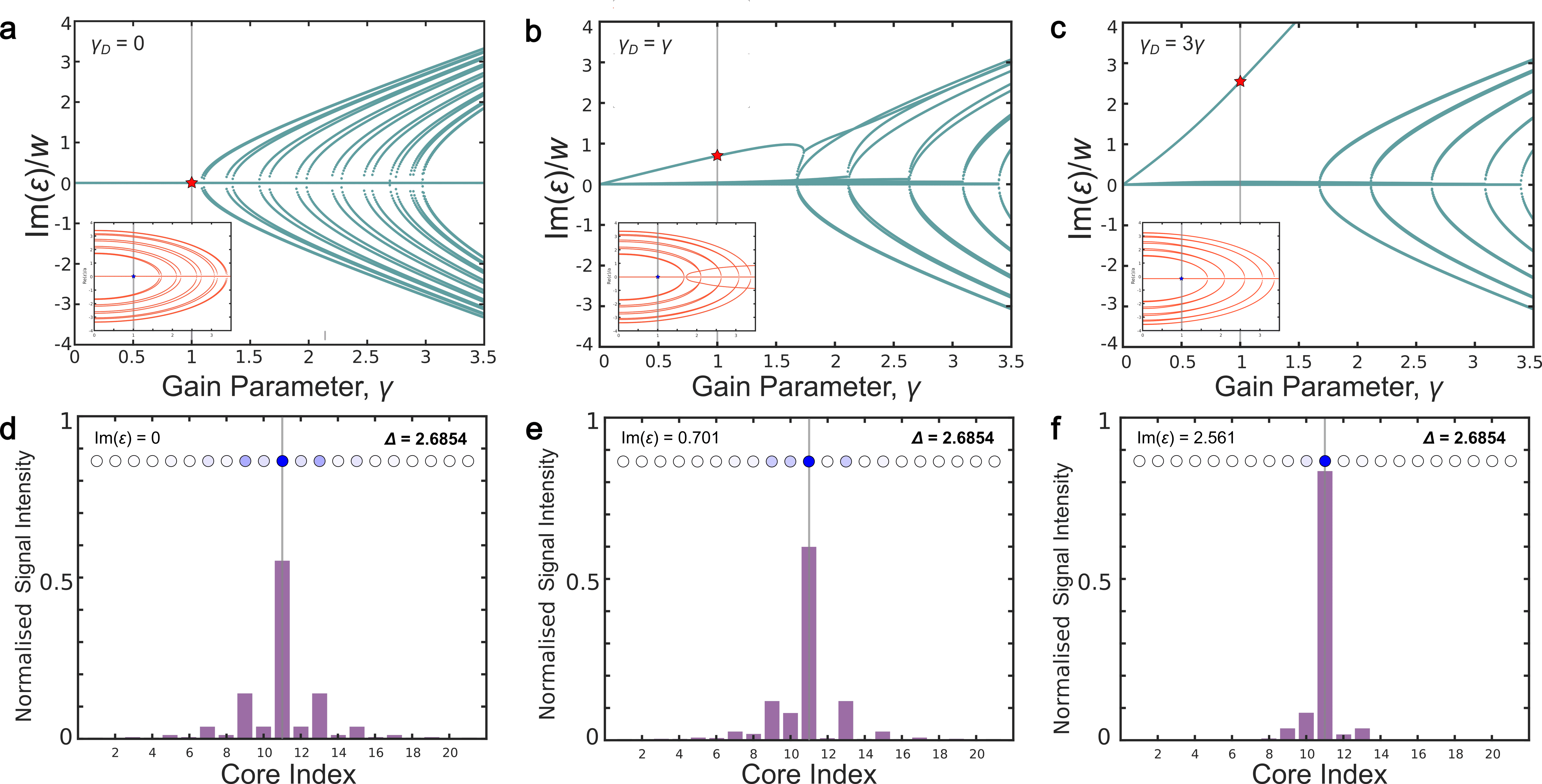} % 70% of text width
    \caption{\textbf{Spectral effects of defect doping}.\quad \textbf{a-c} \quad The imaginary component of the eigenspectrum as a function of the gain parameter, $\gamma$, for defect parameters $\gamma_D = 0$, $\gamma_D = \gamma$, and $\gamma_D = 3\gamma$ respectively. Each is inset with the corresponding real components of the spectra, in orange. \textbf{d-f} \quad Boundary mode profiles with respect to the above spectrum. Plotted mode is marked with red star in \textbf{a-c}, respectively. Each is inscribed with a core-wise heat map, the imaginary eigenvalue, $\mathrm{Im}(\varepsilon)$, and the spectral bandgap, $\Delta$.
    }
    \label{F:2}
\end{figure*}

\subsection{Non-Hermitian Su-Schrieffer-Heeger Lattices}

The Su-Schrieffer-Heeger (SSH) model, originally proposed to describe solitonic excitations in polyacetylene, has become prototypical in exploring topological phases of matter in various platforms, including and not limited to acoustics~\cite{acousticSSH}, solitons~\cite{SolSSH, SolSSH2}, and polaritons~\cite{PolSSH}. In the Hermitian case, SSH lattices include a linear arrangement of sites with alternating strong and weak coupling strengths, $v$ and $w$, respectively. This dimerisation leads to a natural sublattice symmetry and topological band-structure characterised by a non-zero winding number. At interfaces where the integer winding number changes value, for example at the edge of the lattice or at an engineered interface, robust zero-energy boundary states are observed as a consequence of the bulk-boundary correspondence.

In recent years, attention has shifted towards non-Hermitian extensions of the SSH model, in open systems that are naturally dissipative or non-conservative, or where non-Hermiticity is deliberately engineered. Including non-Hermitian terms in the model fundamentally alters the spectral and topological properties, where eigenvalues shift into the complex plane, and the imaginary component signifies an amplification or attenuation of the corresponding eigenvector, depending on the sign. A particularly rich realisation of non-Hermiticity in the SSH model is obtained by incorporating $\mathcal{PT}$-symmetry, which enables the coexistence of topological features and $\mathcal{PT}$-symmetric physics. While the introduction of non-Hermitian terms into an SSH lattice produces non-trivial changes to the topological behaviour, this area is well-researched in terms of the underlying Berry mechanism and homotopy~\cite{Homo1, Homo2, Homo3}. Significantly, the topological properties of the SSH model can be preserved in some non-Hermitian systems, including the reciprocal $\mathcal{PT}$-symmetric chain~\cite{Schomerus, Zhu, Lang, Weimann, Pan, Li,Zhao}.

The $\mathcal{PT}$-symmetric SSH chain provides a minimal yet powerful model for exploring the interplay of non-Hermiticity and topology. First, the dimerised couplings about a central interface support localised topological boundary modes. Second, $\mathcal{PT}$-symmetry controls the stability of these states and gives rise to unique spectral transition about parametric \textit{exceptional points}. The $\mathcal{PT}$-symmetry breaking beyond these exceptional points introduces an interesting regime in which topological boundary modes are selectively amplified while  bulk modes are suppressed. A schematic of this arrangement is shown in Fig.~\ref{F:1}a. The arrangement is comprised of two mirrored SSH chains joined at a central defect (blue). The weak and strong couplings are indicated by single and double dashed lines between cores, respectively, and green/red regions indicate sites of gain and loss. These are arranged in a $\mathcal{PT}$-symmetric fashion such that the gain profile is odd about the central defect, as required. The SSH chain may be coiled into a photonic crystal fibre (PCF), shown in Fig.~\ref{F:1}b, where weak and strong couplings are engineered via alternating diameters of airholes between light guiding regions, inspired by the experimental fabrication of a Hermitian fibre in Ref.~\cite{BathSSH}. An additional small airhole is placed to the left of the central core to ensure all cores have the same local spatial environment. The antisymmetric gain and loss may be implemented by carefully tuning concentrations of rare-earth and transition metal dopants, respectively. Level diagrams of the pump and absorption mechanisms for both types of dopant type are included in Fig.~\ref{F:1}c. Ytterbium ions amplify signals at an operational wavelength of 1050nm via stimulated emission when pumped at 940nm. On the contrary, chromium ions absorb photons at 1050nm, which is dissipated non-radiatively as phononic excitations. By carefully tuning the concentration of each dopant, the gain and loss profile can be incorporated into the design symmetrically. Fig.~\ref{F:1}d shows the output steady-state signal as calculated through a computational propagation simulation, overset with a heat-map distribution of intensity. We note that topological localisation is observed even in the prescence of nonlinear saturable gain. A slight skewing of intensity is also observed as a result of explicit $\mathcal{PT}$-symmetry breaking due to the presence of gain at the central defect site.

\section{\label{S:3}Topological Characteristics of Near-$\mathcal{PT}$ symmetric system}

\subsection{Spectral Behaviour}
Given that the chosen arrangement is $\mathcal{PT}$-symmetric, we expect that the spectrum will undergo spontaneous $\mathcal{PT}$-symmetry breaking at gain values beyond the exceptional point. The resulting spectrum becomes complex, with pairs of modes being amplified and attenuated. The imaginary part of the eigenvalues are plotted as a function of the gain parameter, $\gamma$ in Fig.~\ref{F:2}a. The orange inset shows the real spectrum. Many exceptional points emerge in the imaginary plane while symmetric pairs of modes become degenerate. It must be noted that the topological mode situated at zero energy is not affected by spontaneous $\mathcal{PT}$-symmetry breaking: it retains zero imaginary component and is neither amplified nor attenuated. \\

In order to selectively amplify the topological mode, we use the explicit $\mathcal{PT}$-symmetry breaking in the form of gain on the defect core. This is parametrised by $\gamma_D$. When $\gamma_D > 0$, the gain and loss profile becomes unbalanced, which has non-trivial consequence for the real and imaginary spectra of the system. In Fig.~\ref{F:2}b and Fig.~\ref{F:2}c, the spectra are plotted for a defect gain of $\gamma_D = \gamma$ and $\gamma_D = 3\gamma$, respectively. The transition is characterised by a leapfrogging exceptional points and the eventual selective amplification of the boundary mode, marked therein with a red star. In Fig.~\ref{F:2}c, the red marking highlights an area of interest wherein a single mode is amplified while all other modes remain suppressed. Since this regime is situated before the exceptional point, the spectral gap (see orange inset) remain large, hinting at a sustained topological bandgap despite non-Hermiticity and explicit $\mathcal{PT}$-symmetry breaking. The profile of the central modes marked with a red star are plotted respectively  in Fig.~\ref{F:2}d-f. Each is inset with the magnitude of the imaginary component, $\mathrm{Im}(\varepsilon)$, corresponding to amplification. The spectral bandgap, $\Delta$ is largely independent of defect doping, and we conclude that the topology is preserved despite the introduction of non-Hermiticity by adiabatic continuity. A rightward skew in the mode profile is observed due to the $\mathcal{PT}$-asymmetry introduced by defect doping. As a consequence of the topology, the mode remains localised to the central defect. 

\begin{figure*}[t]
    \includegraphics[width=1\textwidth]{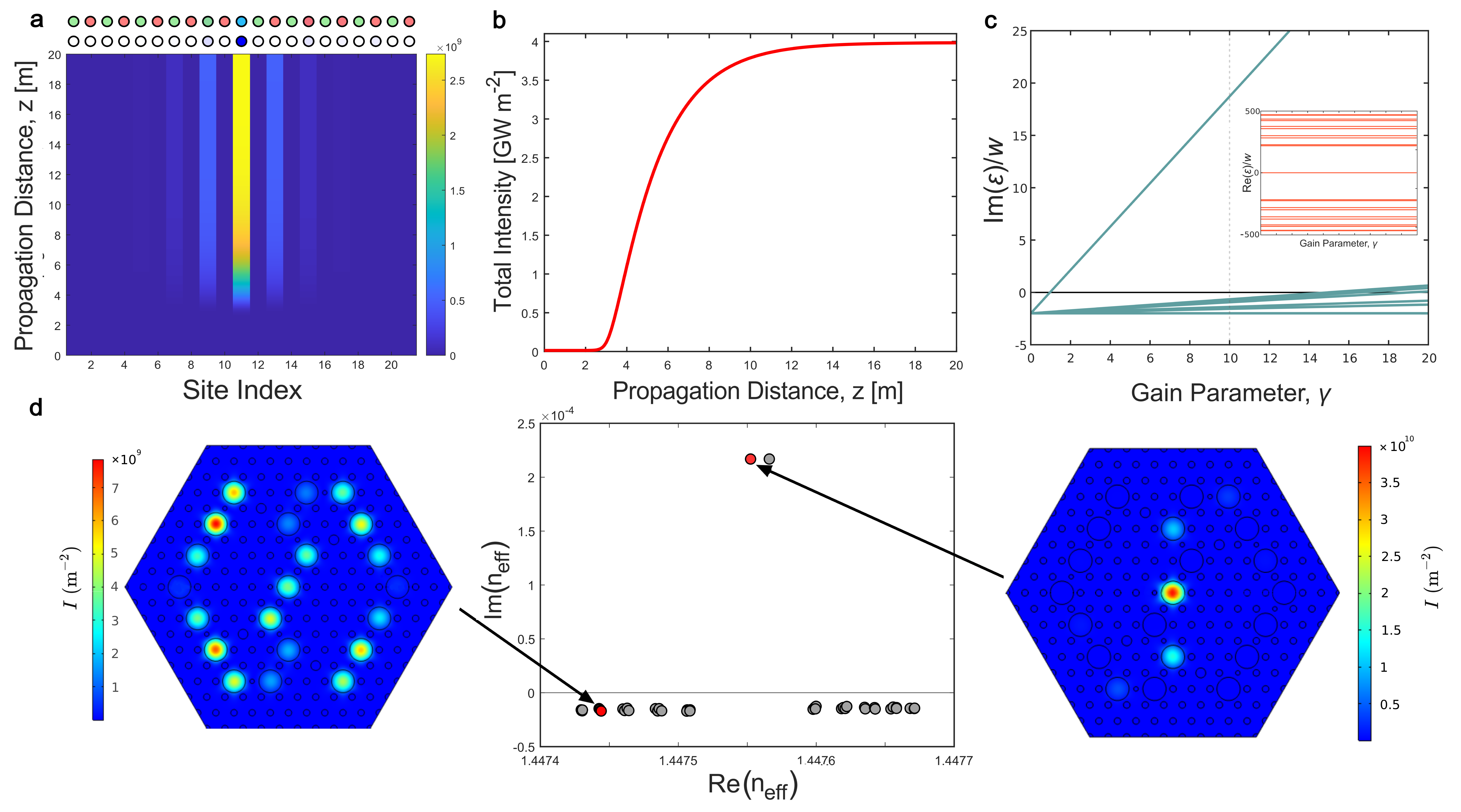} % 70% of text width
    \caption{\textbf{Spectral and dynamic simulations.}
    \textbf{a}\quad Tight-binding propagation simulation displaying local signal evolution along propagation direction.
    \textbf{b}\quad Total power as a function of propagation distance.
    \textbf{c}\quad The imaginary modal spectrum of the proposed design as a function of gain parameter, $\gamma$. Therein, a grey vertical dash marks the parametric regime of the tight-binding model ($\gamma$ = 10). Inset highlights the apparent invariance of the real component of the spectrum under action of the same gain parameter shift.
    \textbf{d}\quad (\textbf{centre}) Complex spectrum of topological fibre laser simulated using finite element method. Two modes are highlighted in red and plotted to either side: \,\ (\textbf{left}) Relative intensity  $I = \frac{|E|^2}{\int |E|^2 dA} $ of a bulk mode. \ \ (\textbf{right}) Relative intensity $I$ of a topological edge mode.}
    \label{F:3}.
\end{figure*}

\section{Practical Consideration and Computational Results}\label{S:4}

\subsection{Design Specifications}\label{S:Spec}

The proposed design is a silica glass photonic crystal fibre whose light-guiding regions are doped with rare-earth and transition metal dopants in order to engineer balanced regions of gain and loss. The gain regions are doped with ytterbium(III) oxide ($\mathrm{Yb}_2\mathrm{O}_3$) at a concentration of $1800\,\mathrm{ppmw}$ (parts per million by weight). The loss regions are doped with chromium(III) oxide ($\mathrm{Cr}_2\mathrm{O}_3$) at a concentration of $55\,\mathrm{ppmw}$. These are tuned to correspond to an equal and balanced gain/loss coefficient of $\gamma = \pm\, 10\, \mathrm{m}^{-1}$, respectively. The central defect gain parameter, $\gamma_D = 3\gamma$, is doped $\mathrm{Yb}_2\mathrm{O}_3$ at $5400\,\mathrm{ppmw}$, yielding an effective gain parameter of $\gamma_D = +\, 30\, \mathrm{m}^{-1}$. The core size is tuned to be single mode for a wavelength of $1050\,\mathrm{nm}$, requiring a pump of $940\,\mathrm{nm}$. The $^{2}\mathrm{F}_{5/2}\, \mapsto\,  ^{2}\mathrm{F}_{7/2}$ electronic transition of ytterbium atoms then facilitates stimulated lasing at the operating wavelength. We assume that the ratio of air hole radii are tuned to $v=350\,\mathrm{m}^{-1}$ and $w=160\,\mathrm{m}^{-1}$, using the parameters from Ref.~\cite{BathSSH}. See Appendix A for these calculations.\\ 

\subsection{Finite Element Analysis}

To more accurately investigate the interplay between topology and non-Hermitian effects in a continuum setting, we simulate the proposed geometry using  COMSOL Multiphysics finite-element software. Since the design is invariant in the direction of propagation, it is sufficient to construct a flat geometry according to Fig.~\ref{F:1}a and perform a normal mode analysis.\\

Computationally, regions of gain and loss are introduced as imaginary coefficients in the material index. These indices are uniformly scaled down by a linear factor to mediate increased numerical instability resulting from large gain/loss values in COMSOL. This can be performed without loss of generality with respect to the spectra in Section III., providing the gain parameter remains in the symmetry-preserving regime below the first exceptional point (i.e., for $\kappa > \gamma)$.\\

Finite element mode analysis of the system produced the  complex spectrum displayed in Fig.~\ref{F:3}d (centre). The supermodes of the system form two photonic bands consisting entirely of trivial bulk modes situated in the lower half-plane, corresponding to a net attenuation. A large bandgap is observed, a common signature in non-Hermitian SSH systems, housing a pair of orthogonal, topologically-protected defect modes, in strong agreement with the tight-binding spectrum shown in Fig.~\ref{F:3}c.

As well as being attenuated by design, the bulk modes exhibit negligible mode overlap with the central core, which inhibits sporadic beating between the topological and trivial supermodes.
Intrinsically, the spectra of $\mathcal{PT}$-symmetric systems are always real, enforced by internal symmetry. Our understanding of topologically-selective amplification is as follows:\\

Hermitian SSH structure is well-studied, and operates via engineered interfaces between coupling patterns. In this case, the chirality of alternating couplings is mirrored about the central core. This mismatch leads to a double weak coupling about the defect and therefore the emergence of a mid-gap boundary mode, via the bulk-boundary correspondence. This has a non-trivial effect on the band structure, and modes are organised into gapped bands, within which the boundary mode is housed. This can be thought of as a continuous deformation of the Hamiltonian, as the lattice is deformed from a uniform arrangement to a dimerised arrangement. As such, the eigenvectors of the system are spatially rearranged such that the topological mode is highly localised at the boundary, while the bulk modes are delocalised. The gain/loss distribution may then be engineered such that the topological mode overlaps a net effective gain, whereas bulk modes overlap a net effective loss.

\subsection{Finite Difference Propagation}

In addition to the spectral properties of the proposed system, the dynamic behaviour of propagating signals is also modelled to understand the modal behaviour in response to nonlinear effects. In practice, gain is achieved with dopants via stimulated emission, which in the photonic picture is a one-to-one process. Therefore, the amount of gain that is achieved will be proportional to the dopant concentration. When a sufficient number of pump photons are incident on the dopant atoms, the electron states will saturate on the highest energy band of Fig.~\ref{F:1}c, and the lowest energy band will be depleted. This describes a \textit{population inversion} and signifies a ceiling for the amount of gain that is achievable for a given concentration. This behaviour is dynamic and may be simulated using coupled mode theory, where the non-Hermitian terms have a saturating nonlinearity~\cite{LaserTheory, NTP, Saturation}:

\begin{figure*}[t]
    \includegraphics[width=1\textwidth]{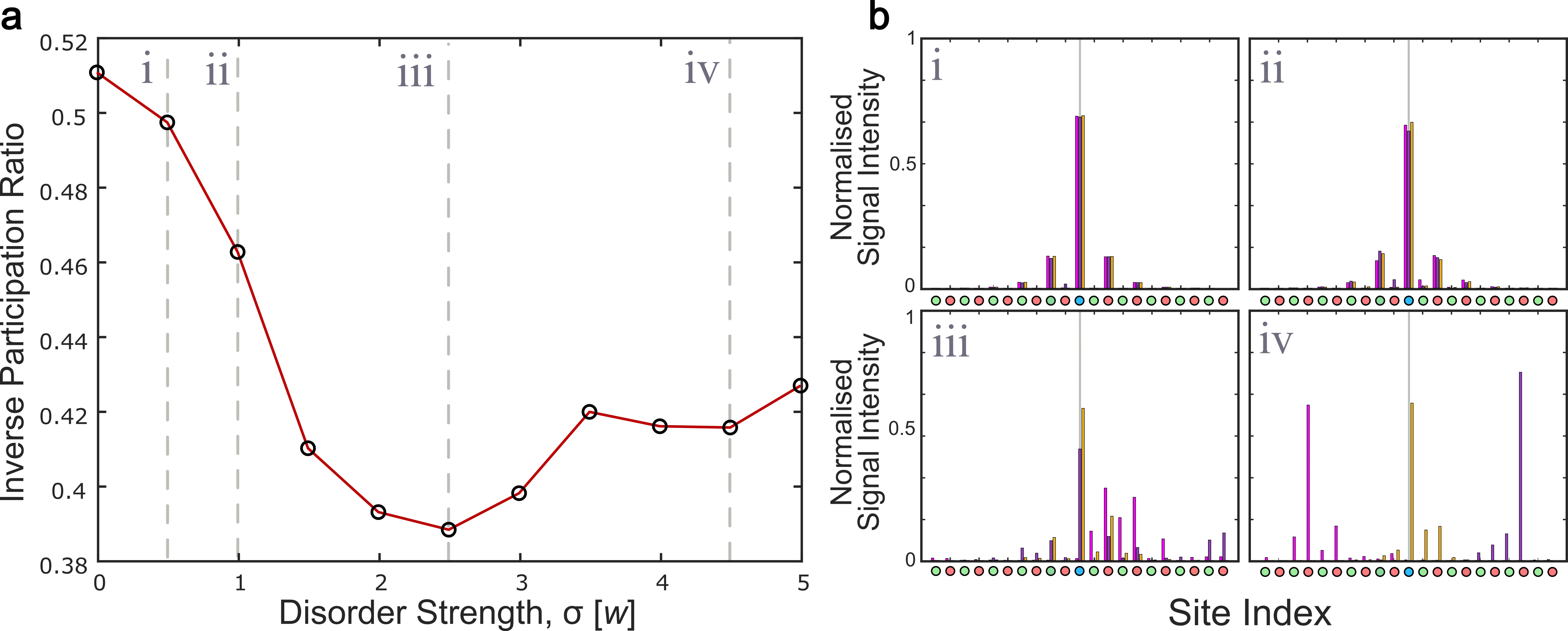} % 70% of text width
    \caption{\textbf{Effects of on-site disorder.}\quad \textbf{a} \quad Inverse participation ratio of output signal after propagating $z = 15\, \mathrm{m}$, as a function of on-site disorder. We use a Gaussian distribution of onsite disorder and vary the standard deviation $\sigma$. Data points are an average of 1000 disorder realisations. \textbf{b}(\textit{i--iv}) show the output mode profile for disorder strengths $\sigma = 0.5w, 1.0w, 2.5w, 4.5w$, respectively. Three instances of random disorder are shown as blue, red and yellow profiles. 
    }
    \label{F:4}
\end{figure*}

\begin{align}
    -i \frac{d \Psi_n}{dz} &= \sum_m C_{nm} \Psi_m -i\alpha \Psi_n+ i\, \gamma_{ss,n} \cdot \left( \frac{1}{1 + \frac{|\Psi_n|^2}{I_{\text{sat}}}} \right) \Psi_n
\end{align}

where $z$ is the longitudinal coordinate pointing along the fibre axis, $C_{nm}$ is the Hermitian coupling matrix, $\alpha$ is the global attenuation coefficient, $I_{sat}$ is the saturation intensity of each core, dependent on the dopant concentration, and $\gamma_{ss,n}$ is the small-signal gain parameter. We see that the initial small-signal parameter is modified in such a way that as the intensity of the signal increases, the gain parameter decreases (and vice versa). A global attenuation coefficient, $\alpha$, is included to account for the natural dissipative qualities of the host medium. Notice that in practice, $\alpha$ is a free parameter, as it corresponds to a global shift in the gain/loss profile of the system, which can be artificially engineered by tuning the active dopant concentrations in the cores. We will see that this serves as a powerful tool for completely suppressing the bulk modes, allowing for single-mode topological lasing. Finite difference techniques are utilised to calculate the field derivate at infinitesimal steps and propagate the signal through a fixed distance, via: 

\begin{align}
    \mathbf{\Psi}(z + \Delta z) = \mathbf{\Psi}(z) + \Delta z \cdot \frac{d\mathbf{\Psi}}{dz},
\end{align}
where $\mathbf{\Psi} = (\Psi_1, \cdots.\Psi_n)$. Results pertaining to these simulations can be found in Fig.~\ref{F:3}. The kymograph
 in Fig.~\ref{F:3}a shows a heatmap of field intensity for each core is plotted as a function of propagation distance (see parameters in Sec.~\ref{S:Spec}). The system reaches a steady-state where light is localised strongly to the central defect. The signal is confined to alternating (odd) sites, a defining feature of topological boundary states in the SSH model. Fig.~\ref{F:3}b shows the total system intensity as a function of propagation distance. We observe that the fibre laser exhibits a smooth, monotonic rise to this steady state, due to the saturating effect of the nonlinear gain. The effect of this offset on the imaginary spectrum is illustrated in Fig.~\ref{F:3}c, where the specific gain parameter is marked with a red line. For these linear parameters, bulk modes are  suppressed and the topological mode is selectively amplified. As seen in the inset of Fig.~\ref{F:3}c, this has a negligible effect on the real spectrum. For these parameters, the coupling strength-to-gain ratio is sufficiently large that non-Hermiticity does not alters the bandgap size, and thus boundary states remain topologically protected. We performed finite-elements analysis for the full two-dimensional profile, and the characteristic lasing mode is plotted in Fig.~\ref{F:3}d (right).

\section{Topological Robustness}

\begin{figure*}[t]
    \includegraphics[width=1\textwidth]{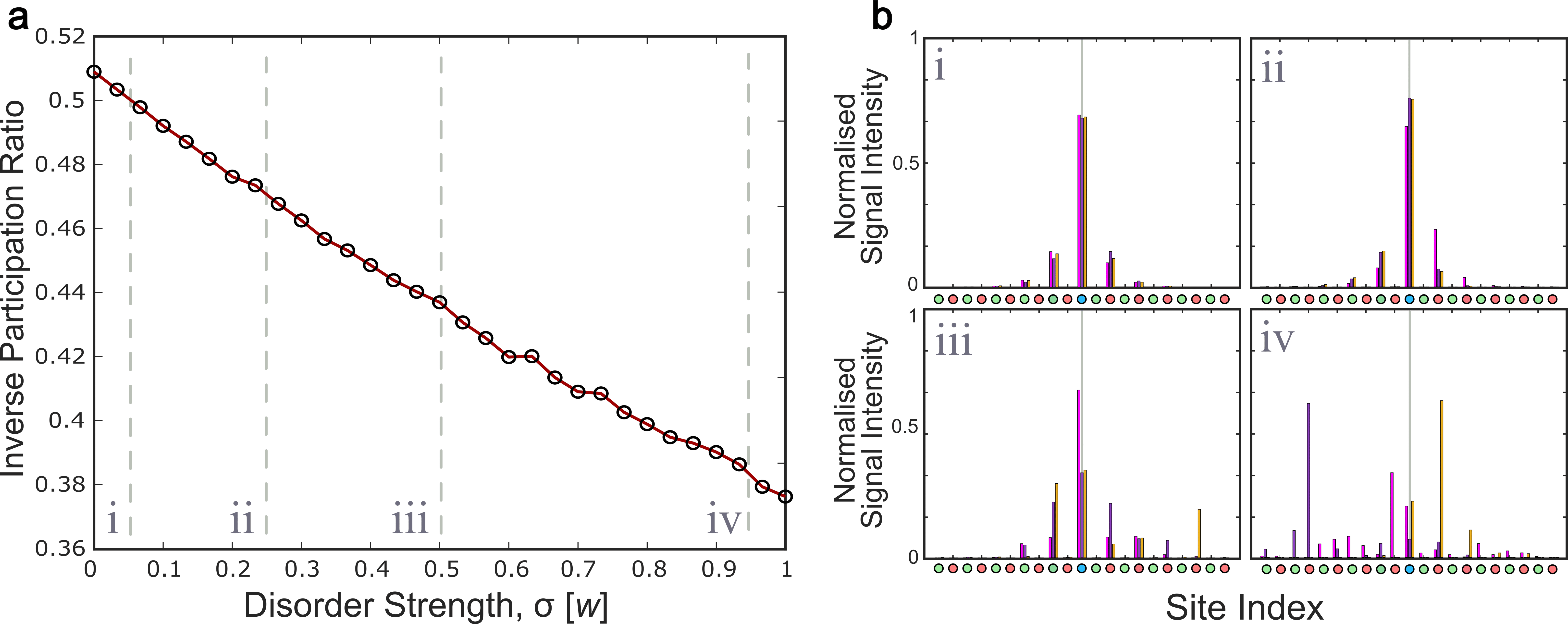} % 70% of text width
    \caption{\textbf{Effect of coupling disorder.}\quad \textbf{a} \quad Inverse participation ratio of output signal after propagating $z = 15\,\mathrm{m}$, as a function of coupling disorder. We use a \emph{uniform} distribution of disorder in the coupling strength disorder (i.e., off-diagonal disorder) and vary the standard deviation $\sigma$. Data points are an average of 1000 disorder realisations. \textbf{b}(\textit{i--iv}) show the output mode profile for $\sigma = 0.05 w, 0.25 w, 0.50 w, 0.95 w$, respectively. Three realisations of disorder are plotted for each $\sigma$, in pink, purple, and yellow. 
    }
    \label{F:5}
\end{figure*}
 
\subsection{Robustness to Onsite Disorder}

The proposed architecture benefits from an increased robustness to fabricated disorder as a consequence of its underlying topological protection. One prominent source of disorder in fabrication arises from the random variation of core size during the fabrication process. This translates into fluctuations of the onsite propagation constants in the coupling matrix. In our simulations, this is modelled by adding Gaussian-distributed random perturbations to the diagonal elements of the coupling matrix, which effectively modifies the refractive index profile of each core.\\

Fig.~\ref{F:4}a shows the results of propagation simulations where the inverse participation ratio (IPR) of the output profile (after propagation distance of 15 meters) is plotted against the strength of on-site disorder. IPR is a measure of spatial localisation, for which low values correspond to a broad, delocalised signal and high values indicate localisation. For discrete modes, it is calculated site-wise for the second and fourth moments,
\begin{align}
    \textrm{IPR}(\mathbf{\Psi}) = \frac{\sum_n \vert\psi_n\vert^4}{(\sum_n\vert\psi_n\vert^2)^2},
\end{align}
where $\mathbf{\Psi}= \sum_n\psi_n$ is the mode amplitude, with $n$th-core signal amplitudes, $\psi_n$. Fig.~\ref{F:4}a demonstrates that the steady state signal remains highly localised in the presence of onsite disorder up to values $\approx 1w$ (where $w$ is the larger nearest-neighbour coupling strength), corroborated by the output profiles plotted in subpanels Fig.~\ref{F:4}b(\textit{i}) and (\textit{ii}). Beyond this value, delocalisation is observed, with IPR reaching a minimum at $\approx 2.5w$, see output profile in Fig.~\ref{F:4}b(\textit{iii}).  For disorder strength beyond $2.5w$, the signal relocalises due to Anderson localisation effects. The mode profile for a Gaussian standard deviation $\sigma = 4.5w$ is plotted in (\textit{iv}). In some instances, nearest neighbour coupling is disrupted and the signal is confined to a single gain core, which is then highly amplified. This is characterised by a plethora a high peaks. This effect of onsite disorder resilience is surprising given that the SSH winding-number invariant is not typically robust to on-diagonal disorder and indicates that the nonlinear non-Hermitian chain offers topological robustness beyond that of the classical SSH chain.

\subsection{Robustness to Coupling Disorder}

Another potential source of disorder during fabrication is the introduction of coupling disorder due to shifting in the relative positions of the cores. As with the previous section, this disorder can be modelled as random changes to the coupling coefficients in the coupling matrix. This is implemented by adding a uniform distribution of random numbers to the off-diagonal entries of the (Hermitian) coupling matrix. The inverse participation ratio after finite-difference propagation is then analysed to quantify topological robustness.  In Fig.~\ref{F:5}a, the output IPR for each disorder strength is calculated in 1000 instances, averaged, and plotted. Fig.~ \ref{F:5}a(\textit{i-iv}) show the output mode profiles for $\sigma = 0.05w, 0.25w, 0.5w$ and $1.0w$, respectively. We observe that the IPR follows an approximately linear delocalisation as a function of disorder strength. 
A strong topological confinement is observed up to a disorder strength of $\sigma = 0.25w$, corroborated by the profile instances in Fig.~\ref{F:5}b(\textit{i-ii}). In these cases, the signal remains localised to the defect core, with peripheral amplitudes situated on a single sublattice, indicative of the topological edge states routinely observed at SSH boundaries.

\section{Conclusion}

We have presented a design for a topological fibre laser based on a non-Hermitian Su-Schrieffer-Heeger chain. We used a design based on a previously fabricated photonic crystal fibre, and modelled the effects of gain and loss due to dopants. By employing explicit $\mathcal{PT}$-symmetry breaking at the topological defect, we show that the boundary mode can be selectively amplified, leading to localization beyond the linear SSH chain, whereas the bulk modes are suppressed. We have confirmed our coupled-oscillator model using realistic finite-difference propagation and finite-element normal-mode simulations. We then demonstrate that the lasing mode retains strong localisation in the presence of saturable nonlinearity and is robust against both on-site and coupling disorder. Our design accounts for realistic parameters and standard fibre fabrication techniques, demonstrating a feasible route to disorder-resilient fibre lasers. Our results offer a general route for combining non-Hermitian physics with nonlinearity and topological protection in fibre systems. Further research would include extending the topological model to other topological invariants, with the goal of creating robust multicore lasing modes that are coherent at the output, with implications in high-power fibre lasers. 

\begin{acknowledgments}
We wish to acknowledge our funding agency, the insightful conversations with Luke Pimlott and Pablo Reiser Ramirez of University of Bath.
This work is supported by the Air Force Office of Scientific Research under award number FA865522-1-7028. A.S.~acknowledges the support of the Royal Society under grant No.~RGS/R2/202135.

\end{acknowledgments}

\appendix

\section{Derivation of Doping Concentration} 

To fabricate the fibre, we hypothesize the following approach: Yb-doped, Cr-doped and two hollow rods of varying airholes. The preform is stacked by hand to create a macroscopic arrangement of elements replicating the design in Fig.~\ref{F:1}a and inserted into a silica outer jacket tube. This preform, which is a few centimetres in diameter, is then fed through a furnace and drawn into smaller preforms known as canes which are a few millimetres in diameter. Finally, the canes are jacketed again and drawn to fibre of $100-200\,\mu \mathrm{m} $ diameter while pressurising the holes to keep them open, resulting in a fibre with structure to support the desired guidance properties.

In order to achieve a balanced gain and loss of $\gamma = 10\,\mathrm{m}^{-1}$, the dopant concentration in each type of core has to be finely tuned. Via \cite{TransitionSilica}, and Table I therein, silica glass doped with $\mathrm{Cr}_2\mathrm{O}_3$ may be prepared in vapour phase via flame hydrolysis, with a  practical absorbability of $0.8\,\mathrm{dB}\,\mathrm{km}^{-1}\,\mathrm{ppbw}^{-1}$ (part per billion by weight) at 1050nm. Aiming for an attenuation of $\gamma = -10\,\mathrm{m}^{-1} = -4343 \,\mathrm{dB}\,\mathrm{km}^{-1}$, the concentration is then given by
\begin{align}
    C_{[\mathrm{Cr}_2\mathrm{O}_3]} = 4343\div 0.8 = 54287\,\mathrm{ppbw} \approx 54.3\,\mathrm{ppmw}.
\end{align}

This is within the range of sensible doping concentrations for chromium, which has been reported in concentrations as high as $1200\,\mathrm{ppmw}$~\cite{HighCr} in silica glass. In gain cores we are aiming for a balanced amplification of $\gamma = +10\,\mathrm{m}^{-1}$. The small signal gain coefficient, $\gamma_{ss}$, (in the best case; full population inversion), may be approximated by, 
\begin{align}
    g \approx \sigma_e C_{[{Yb}^{3+}]}
\end{align}

where $\sigma_e$ is the emission cross-section of $\mathrm{Yb}^{3+}$. Spectroscopic studies of $\mathrm{Yb}^{3+}$-doping in silica glass determine a emission cross-section of $0.7\times 10^{-24}\,\mathrm{m}^2$~\cite{YbSilica}. Eq.(A2) yields an $\mathrm{Yb}^{3+}$ ion concentration of $C_{[{Yb}^{3+}]} = 1.4 \times 10^{25}\,\mathrm{m}^{-3} \approx 1800\,\mathrm{ppmw}$ in silica glass.\newline

We note that all codes and data used for both the finite-difference and finite-element simulations may be found on the cited Zenodo repository.~\cite{Zenodo}.

\bibliography{apssamp}

\end{document}